\documentclass[10pt,conference]{IEEEtran}
\IEEEoverridecommandlockouts
\usepackage{cite}
\usepackage{amsmath,amssymb,amsfonts}
\usepackage{graphicx}
\usepackage{textcomp}
\usepackage{xcolor}
\usepackage{algorithm,algorithmic}
\usepackage{subcaption}
\usepackage{enumitem}
\usepackage{float}
\usepackage{amsmath}
\usepackage{color,soul}
\usepackage{multirow}
\usepackage{cleveref}
\usepackage{stfloats}
\crefname{figure}{fig.}{figures}
\Crefname{figure}{Fig.}{Figures}
\crefname{equation}{eq.}{eqs.}
\Crefname{equation}{Eq.}{Eqs.}
\def\BibTeX{{\rm B\kern-.05em{\sc i\kern-.025em b}\kern-.08em
    T\kern-.1667em\lower.7ex\hbox{E}\kern-.125emX}}
\begin{document}

\title{Variational Quantum Linear Solver for Simulating Quantum Transport in Nanoscale Semiconductor Devices \\

\thanks{This work was supported by NSF Grants \#2007200 and \#2142552.}
}

\author{\IEEEauthorblockN{Qimao Yang\IEEEauthorrefmark{1} and Jing Guo\IEEEauthorrefmark{2}}
\IEEEauthorblockA{\IEEEauthorrefmark{1}\IEEEauthorrefmark{2}Department of Electrical and Computer Engineering,\\
University of Florida, Gainesville, FL, 32611 USA\\
\IEEEauthorrefmark{1}qimao.yang@ufl.edu,
\IEEEauthorrefmark{2}guoj@ufl.edu}}

\maketitle
\begin{abstract}
This work develops simulation methods that enable the application of the variational quantum linear solver (VQLS) to simulate quantum transport in nanoscale semiconductor devices. Most previous work on VQLS applications in semiconductor device simulations focuses on solving the Poisson equation, where the coefficient matrix of the sparse linear system is real and symmetric. Solving the quantum transport equation, however, leads to coefficient matrices that are complex and non-symmetric. This work addresses the challenges of applying VQLS to quantum transport simulations. We propose new forms of cost functions to solve complex and non-symmetric linear systems with faster computing speed. We further develop efficient decomposition methods for cost function evaluation, which target reducing the quantum circuit complexity and improving noise robustness when solving the quantum transport equation using the non-equilibrium Green's function method. While classical computation faces the challenge of the "curse of dimensionality" as the spatial-energy numerical grid dimensions grow, the proposed quantum-computing-based method scales logarithmically with the grid size, which offers a promising opportunity for addressing the computational challenges of solving quantum transport in semiconductor devices.
\end{abstract}


\section{Introduction}
Modeling, simulation, and design play a critical role in semiconductor technologies. As advanced transistors scale down to the nanometer-scale regime, quantum transport simulations are essential for capturing quantum effects important for nanoscale semiconductor devices\cite{datta_quantum_2005}. Semiconductor device simulation often involves solving partial differential equations from physics such as the Poisson equation for electrostatic potential and the Non-Equilibrium Green's Function (NEGF) method for quantum transport calculations\cite{jiangBoundaryTreatmentsNonequilibrium2008, lundstromFundamentalsCarrierTransport2000}. Compared to the Poisson equation, whose discretized form often involves a large symmetric real system of linear equations, the quantum transport equation is more challenging to solve. It has a higher dimensionality and involves complex, non-symmetric system equations\cite{datta_quantum_2005, paulssonNonEquilibriumGreens2006}. Simulating quantum transport in classical computers is computationally intensive and often slow, which limits semiconductor device design capabilities.

Recent advances in quantum computing have opened new possibilities for solving computationally intensive problems in scientific and engineering domains. Among various quantum algorithms, the Variational Quantum Linear Solver (VQLS)\cite{bravo2023variational} has emerged as a promising approach to solving linear systems of equations on near-term quantum computers\cite{trahanVariationalQuantumLinear2023, gnanasekaranEfficientVariationalQuantum2024}. It represents a hybrid quantum-classical approach that can potentially address the computational bottleneck in quantum transport simulations through its logarithmic scaling with problem size, compared to the exponential scaling faced by classical methods.

However, applying VQLS to semiconductor device modeling—particularly for quantum transport simulations—faces several fundamental challenges. Previous work on VQLS has primarily focused on symmetric and real-numbered systems\cite{liu2021variational, liVariationalQuantumAlgorithms2023a}, with various cost functions proposed in the literature. Global and local cost functions\cite{bravo2023variational} often suffer from poor convergence. Alternatively, the energy-based cost function shows improved convergence\cite{sato2021variational}, but is limited to solving real symmetric linear systems. Solving quantum transport requires treating a complex-numbered, non-symmetric system of equations, creating a significant gap between current VQLS capabilities and the requirements of semiconductor device simulation.

This work bridges this gap through two main contributions. First, we develop a novel hybrid cost function design that effectively handles both non-symmetric matrices and complex-numbered systems while improving convergence behavior. Our approach combines normalized residual minimization with the theoretical rigor of local cost function, providing a more robust framework for semiconductor device simulations. In addition, we prove that this new cost function formulation can be efficiently decomposed without increasing measurement requirements compared to existing approaches. Second, we propose an efficient ansatz construction that enhances the representation capabilities for quantum transport problems.

Our approach enables the application of quantum computing to realistic semiconductor device simulations, potentially offering logarithmic scaling with both spatial and energy grid dimensions. This represents a significant advantage over classical methods, which scale exponentially with problem size, and offers a path toward high-fidelity quantum transport simulations that would otherwise be computationally prohibitive.

The rest of this paper is organized as follows: Section \ref{sec: relatedwork} reviews related work on VQLS and its applications in scientific computing. Section \ref{sec: approach} presents our theoretical framework, including the hybrid cost function design and its decomposition method, followed by our ansatz construction. Section \ref{sec: results} demonstrates the effectiveness of our approach through numerical experiments. Finally, Section \ref{sec: conclusion} concludes with a discussion of implications and future directions.

\section{Related Work} \label{sec: relatedwork}
Quantum algorithms for solving linear systems have seen significant development over the past decade. The Harrow-Hassidim-Lloyd (HHL) algorithm\cite{harrowQuantumAlgorithmLinear2009} first demonstrated the potential quantum advantage for linear systems, achieving exponential speedup over classical methods for sparse matrices. However, HHL and its improvements\cite{doi:10.1137/16M1087072} require fault-tolerant quantum computers that are not yet available\cite{en17051039}. This has motivated the development of variational approaches suitable for near-term quantum devices\cite{tillyVariationalQuantumEigensolver2022, vaquero-sabaterPhysicallyMotivatedImprovements2024}.

The Variational Quantum Linear Solver (VQLS) was introduced as a hybrid quantum-classical algorithm for solving linear systems on Noisy Intermediate-Scale Quantum (NISQ) devices\cite{bravo2023variational}. In VQLS, $|\psi\rangle$ represents the variational quantum state that approximates the solution $|x\rangle$ to the linear system $A|x\rangle = |b\rangle$. This state is prepared using a parameterized quantum circuit (ansatz) that generates $|\psi(\theta)\rangle=U(\theta)|0\rangle$, where $U(\theta)$ is a parameterized unitary operation and $|0\rangle$ is the initial state. The success of VQLS largely depends on the design of appropriate cost functions to guide the optimization process. Bravo-Prieto et al. proposed a global cost function\cite{bravo2023variational} that aims to directly maximize the overlap between $|A|\psi\rangle$ and $|b\rangle$:
{
\setlength{\abovedisplayskip}{4pt}
\setlength{\belowdisplayskip}{4pt}
\begin{align}
C_{G} = 1 - \frac{|\langle b|A|\psi\rangle|^2}{\langle\psi|A^\dagger A|\psi\rangle}. \label{eq:globalcost}
\end{align}
}

They also introduced a local cost function that offers potential advantages in mitigating barren plateaus by focusing on local discrepancies rather than global quantities:
{
\setlength{\abovedisplayskip}{4pt}
\setlength{\belowdisplayskip}{4pt}
\begin{align}
C_{L} = 1 - \frac{\langle\psi|A^\dagger(U_b(I-\frac{1}{n}\sum_{j=0}^{n-1}|0_j\rangle\langle 0_j|\otimes I_{\bar{j}})U_b^\dagger)A|\psi\rangle}{\langle\psi|A^\dagger A|\psi\rangle},  \label{eq:localcost}
\end{align}
}
where $U_b$ is the unitary operator that prepares state $|b\rangle$ ($U_b|0\rangle = |b\rangle$), $| 0_j \rangle$ denotes the zero state on qubit $j$, and $I_{\bar{j}}$ denotes the identity on all qubits except qubit $j$.

Sato et al. introduced an approach based on the concept of minimal potential energy for solving Poisson equations\cite{sato2021variational}, formulating the cost function as:
{
\setlength{\abovedisplayskip}{4pt}
\setlength{\belowdisplayskip}{4pt}
\begin{align}
C_{E} = -\frac{1}{2}\frac{(\langle b|\psi\rangle + \langle \psi|b \rangle)^2}{\langle\psi|A|\psi\rangle}. \label{eq:energycost}
\end{align}
}

This energy-based formulation offers better convergence properties but requires a positive definiteness operator $A$, limiting its applicability to real-valued, symmetric systems.

Efficient matrix decomposition is another critical aspect of implementing VQLS on quantum hardware. Liu et al. proposed a decomposition scheme for the Poisson equation's coefficient matrix under simple operators $\{I, \sigma_+, \sigma_-\}$\cite{liu2021variational}. Their method requires $O(\text{poly }n)$ terms for decomposition but depends on measurement circuits with ancillary qubits, where $n$ is the qubit number. Sato et al. presented an alternative decomposition that uses a constant number of terms independent of the problem size $O(1)$\cite{sato2021variational}. Choi and Ryu recently developed a strategy that balances the number of required quantum circuits and their noise robustness\cite{choi2024variational}, particularly for multi-dimensional Poisson equations with mixed boundary conditions.

The choice of ansatz architecture plays a crucial role in VQLS performance\cite{qinReviewAnsatzDesigning2023}. Hardware-efficient ansatz\cite{kandala2017hardware} use native gates to minimize circuit depth, but often lack expressibility for complex solution spaces. Problem-inspired ansatz\cite{barkoutsosQuantumAlgorithmsElectronic2018, meiromPANSATZPulsebasedAnsatz2023} that incorporate physical symmetries and constraints have shown promise for specific applications. Recent work has explored adaptive\cite{grimsleyAdaptiveVariationalAlgorithm2019} and dynamic approaches\cite{patil_variational_2022} that modify the ansatz structure during optimization to escape local minima.

Quantum transport simulations using the Non-Equilibrium Green's Function (NEGF) formalism represent a particularly relevant application domain for VQLS, as they involve solving large linear systems arising from quantum transport equations. While classical methods for NEGF are well-established, they face computational challenges for realistic device sizes. Previous quantum approaches have primarily focused on simpler electronic structure problems or Poisson equations for electrostatic potential calculations, leaving the full NEGF treatment with complex, non-symmetric matrices as an open challenge.

This work aims to bridge this gap by developing a VQLS approach specifically tailored for NEGF calculations, building on recent advances while addressing the unique challenges posed by quantum transport problems.

\section{Approach} \label{sec: approach}


\subsection{Quantum Transport Equation for Semiconductor Device Simulation}
In modern semiconductor devices, such as nanoscale transistors, magnetic tunnel junctions (MTJs)\cite{duElectricalManipulationDetection2023}, and ferroelectric tunnel junctions (FTJs)\cite{shekhawatDataRetentionLow2020}, quantum transport phenomena play a fundamentally important role for understanding device characteristics and enabling device operation. Simulating these quantum phenomena accurately is crucial for predicting device behavior and guiding design optimization.

The Non-Equilibrium Green's Function (NEGF) formalism has emerged as a powerful and unified computational approach for modeling quantum transport in nanoscale devices. It has been successfully applied to a wide variety of nanoelectronic systems, including silicon nanotransistors\cite{shan_electronic_2011}, graphene transistors\cite{chen_achieving_2015}, carbon nanotube transistors\cite{aravind_simulation_2016}, two-dimensional-semiconductor-based transistors\cite{yangComputationalScreeningMultiscale2022}, magnetic tunnel junctions, ferroelectric tunnel junctions, and superconductor devices.

At the core of the NEGF approach is the calculation of the retarded Green's function:
{
\setlength{\abovedisplayskip}{4pt}
\setlength{\belowdisplayskip}{4pt}
\begin{align}
    G(E)=[(E+i0^+)I-H-\Sigma_1-\Sigma_2-\Sigma_S]^{-1}.
\end{align}
}

Here, $E$ represents the energy level of the injection wave, $i0^+$ denotes an infinitesimally small positive imaginary shift, and $I$ is the identity matrix. The device Hamiltonian $H=H_0+E_{sub}$ comprises the bare Hamiltonian $H_0$ and the subband energy profile $E_{sub}$. The terms $\Sigma_1$ and $\Sigma_2$ represent self-energies from the source and drain contacts, while $\Sigma_S$ accounts for scattering processes through the self-consistent Born approximation. Once the Green's function is determined, key physical quantities such as local density of states, transmission probability, charge density, and current can be derived.

\begin{figure*}[htb!]
    \centering
    \includegraphics[width=0.85\linewidth]{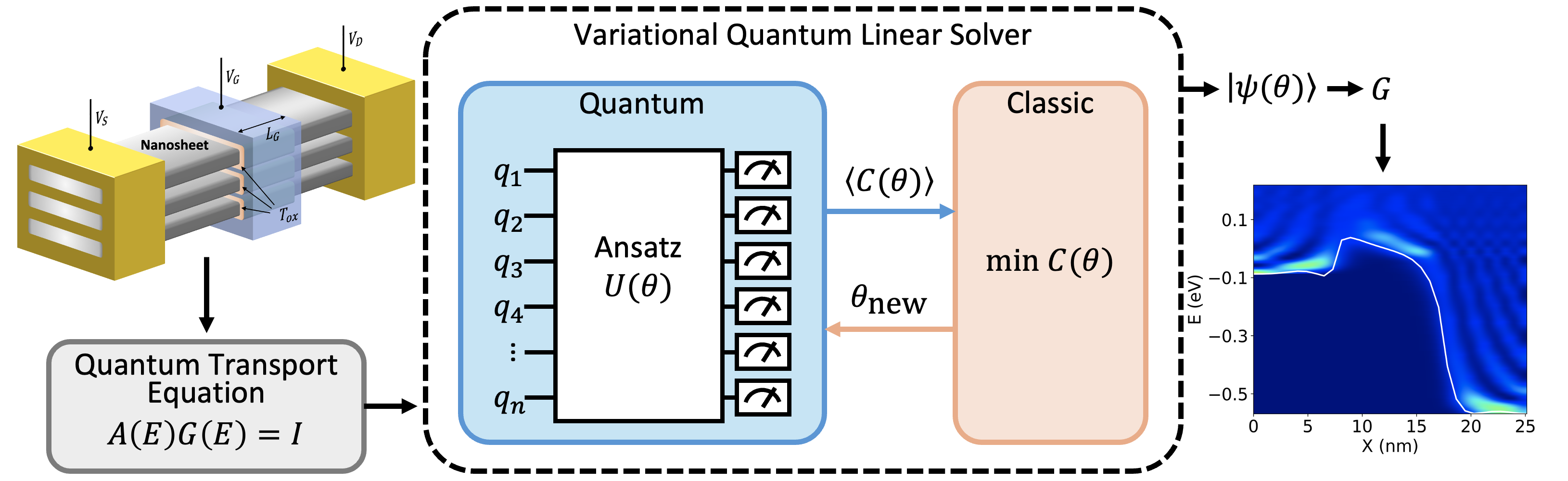}
    \caption{Hybrid quantum-classical framework for simulating quantum transport in nanosheet transistors using a VQLS. The NEGF-based transport equation is reformulated as a linear system, where the Green's function is solved by preparing a parameterized quantum state $|\psi(\theta)\rangle$ via an ansatz circuit. A classical optimizer iteratively updates the parameters $\theta$ to minimize a cost function $C(\theta)$. Upon convergence, the optimized state encodes the Green's function, enabling reconstruction of quantum transport observables.}
    \label{fig:vqNEGF_diag}
    \vspace{-15pt}
\end{figure*}

A key computational challenge of applying NEGF simulations to realistic device structures is its computational intensity. Various approximations and sacrifices in describing realistic device structures and physical effects have to be made to make the simulation computationally tractable on state-of-the-art classical computers. Even with advanced classical algorithms such as the recursive method, the computational complexity scales as $O(N_r N_E)$ for a one-dimensional tight-binding chain with $N_r$ real space grid points and $N_E$ energy grid points.

The NEGF formalism can be reformulated as a linear system:
{
\setlength{\abovedisplayskip}{4pt}
\setlength{\belowdisplayskip}{4pt}
\begin{align}
    A(E)G(E) = I,
\end{align}
}
where $A(E) = (E + i0^+)I - H - \Sigma_1 - \Sigma_2 - \Sigma_S$. From this, each column $G_i(E)$ of the Green's function satisfies:
{
\setlength{\abovedisplayskip}{4pt}
\setlength{\belowdisplayskip}{4pt}
\begin{align}
    A(E)G_i(E) = b_i,
\end{align}
}
with $b_i$ being the $i$-th column of the identity matrix. This formulation transforms the NEGF problem into solving multiple independent linear systems, making it potentially addressable using VQLS as illustrated in Figure \ref{fig:vqNEGF_diag}.



\subsection{Overview Design of Variational Quantum Linear Solver}
VQLS transforms the task of solving a linear system $A(E)G_i(E) = b_i$ into an optimization problem. The solution vector is represented as a parameterized quantum state $|\psi(\theta)\rangle = U(\theta)|0\rangle$, where $U(\theta)$ is a quantum circuit (ansatz) with tunable parameters $\theta$. A classical optimizer then adjusts these parameters to minimize a cost function that quantifies how well the state $|\psi(\theta)\rangle$ approximates the true solution.

The general framework of VQLS for solving the NEGF problem proceeds as follows:
\begin{algorithm}
\small
 \caption{Variational Quantum Linear Solver (VQLS) for NEGF Simulation}
 \begin{algorithmic}[1]
 \renewcommand{\algorithmicrequire}{\textbf{Input:}}
 \renewcommand{\algorithmicensure}{\textbf{Output:}}
 \REQUIRE Hamiltonian $H$, energy $E$, self-energies $\Sigma_1$, $\Sigma_2$, $\Sigma_S$, right-hand side vector $b_i$
 \ENSURE Green's function column $G_i(E)$
 
 \STATE Discretize the problem into matrix equation: $A(E)G_i(E)=b_i$

 \STATE Decompose cost function into a linear combination of Pauli strings
 
 \STATE Initialize parameters $\theta$ for ansatz circuit $U(\theta)$
 
 \FOR{$i = 0$ to max\_iterations}
    \STATE Use quantum subroutines (Hadamard test, swap test, etc.) to measure cost function components
    
    \STATE Compute cost function value $C(\theta)$ using classical processor
    
    \IF{convergence criteria satisfied}
        \STATE $\theta_\text{opt} \gets \theta$
        \STATE \textbf{break}
    \ENDIF
    
    \STATE Update parameters $\theta \gets \text{Optimizer}(\theta, C(\theta))$ to minimize $C(\theta)$
 \ENDFOR
 
 \STATE Compute optimal scaling factor $k^*$ based on optimized state $|\psi(\theta_\text{opt})\rangle$
 
 \RETURN $G_i(E) = k^* \cdot |\psi(\theta_\text{opt})\rangle = k^* \cdot U(\theta_\text{opt})|0\rangle$
 
 \renewcommand{\algorithmicensure}{\textbf{Note:}}
 \ENSURE Computation across multiple energy levels and columns of the Green's function can be quantum parallelized with logarithmic qubit scaling
 \end{algorithmic} 
\end{algorithm}
\vspace{-8pt}

While this framework holds promise for achieving the logarithmic scaling advantage mentioned previously, implementing it effectively for NEGF problems requires addressing the challenges identified earlier. The complex, non-symmetric nature of matrix $A$ necessitates a cost function that can efficiently guide the optimization process without requiring symmetry or positive definiteness.

The success of VQLS for quantum transport simulations depends on several critical design choices. First, the cost function must be carefully designed to handle complex, non-symmetric matrices while maintaining good convergence properties. Second, efficient decomposition strategies must be employed to evaluate the cost function with minimal quantum resources. Third, the ansatz must be sufficiently expressive to represent the solution while remaining trainable on NISQ devices. The following sections address each of these critical aspects in detail.




\subsection{Cost Function Design}
A critical component of the VQLS framework is the cost function $C(\theta)$ that guides the classical optimization process\cite{cerezoCostFunctionDependent2021}. For the NEGF problem with its complex, non-symmetric coefficient matrices $A(E)$, existing cost functions have significant limitations. Global and local cost functions (\cref{eq:globalcost} and \cref{eq:localcost}) are theoretically rigorous but often face practical convergence challenges. Energy-based cost functions (\cref{eq:energycost}) offer better convergence for certain problems but require A to be positive-definite, making them unsuitable for NEGF formalism where A(E) is inherently complex and non-symmetric.


These limitations motivate the development of cost functions specifically tailored to address the challenges of applying VQLS to NEGF simulations. For simplicity in notation, we denote $b = b_i$ (which is naturally normalized) and the target normalized solution state as $|\psi(\theta)\rangle = \frac{G_i(E)}{\|G_i(E)\|_2}$. Our goal is to find parameters $\theta$ such that $A|\psi(\theta)\rangle$ approximates $|b\rangle$.

\subsubsection{Normalized-Residual Cost Function}
To overcome the limitations of existing cost functions when applied to complex, non-symmetric matrices, we propose a normalized-residual cost function based on minimizing the squared norm of the residual $Ax - b$:
{
\setlength{\abovedisplayskip}{4pt}
\setlength{\belowdisplayskip}{4pt}
\begin{align}
     C'_{NR}(x) &= (Ax-b)^\dagger(Ax-b) \nonumber \\
                    &= x^\dagger A^\dagger A x - x^\dagger A^\dagger b - b^\dagger A x + b^\dagger b.
\end{align}
}

Since $b^\dagger b = 1$, this term becomes a constant offset and can be omitted during minimization. To represent the solution as a quantum state, we parameterize the unnormalized solution vector as $|x\rangle = k|\psi(\theta)\rangle$, where $|\psi(\theta)\rangle$ is the normalized parameterized state prepared by the ansatz and $k$ is a real-valued scaling factor. Substituting this into the expression above (and dropping the constant $b^\dagger b$ term), our cost function becomes dependent on both $k$ and $\theta$:
{
\setlength{\abovedisplayskip}{4pt}
\setlength{\belowdisplayskip}{4pt}
\begin{align}
    C_{NR}(k, \theta) &= k^2\langle \psi(\theta)|A^\dagger A|\psi(\theta)\rangle \nonumber \\
    &- k(\langle b|A|\psi(\theta)\rangle + \langle\psi(\theta)|A^\dagger|b\rangle).
\end{align}
}

We minimize this cost function by optimizing $k$ and $\theta$ iteratively. For a given $\theta$, the optimal scaling factor $k^*(\theta)$ can be determined analytically:
{
\setlength{\abovedisplayskip}{4pt}
\setlength{\belowdisplayskip}{4pt}
\begin{align}
    \frac{\partial C_{NR}(k, \theta)}{\partial k} = 0 \implies k^*(\theta) = \frac{\langle b|A|\psi(\theta)\rangle + \langle\psi(\theta)|A^\dagger|b\rangle}{2\langle\psi(\theta)|A^\dagger A|\psi(\theta)\rangle}
\end{align}
}

Substituting this optimal $k^*(\theta)$ back into the cost function yields a function solely dependent on the ansatz parameters $\theta$:
{
\setlength{\abovedisplayskip}{4pt}
\setlength{\belowdisplayskip}{4pt}
\begin{align}
    C_{NR}(\theta) =& -\frac{(\langle b|A|\psi(\theta)\rangle + \langle\psi(\theta)|A^\dagger|b\rangle)^2}{4\langle\psi(\theta)|A^\dagger A|\psi(\theta)\rangle} \nonumber \\  
    =& -\frac{\Re(\langle b|A|\psi(\theta)\rangle)^2}{\langle\psi(\theta)|A^\dagger A|\psi(\theta)\rangle}, \label{eq:cnr_theta}
\end{align}
}
where $\Re$ denotes the real part. This formulation naturally extends to complex matrices and does not require $A$ to be symmetric or positive definite, making it suitable for NEGF simulations.

\subsubsection{Hybrid Cost Function}
While the normalized-residual cost function (Eq.~\ref{eq:cnr_theta}) appropriately handles complex matrices, we observed that convergence can be further enhanced by incorporating aspects of the local cost function. We introduce a hybrid cost function:
{
\setlength{\abovedisplayskip}{4pt}
\setlength{\belowdisplayskip}{4pt}
\begin{align}
    C(\theta) = \alpha \cdot C_{NR}(\theta) + (1-\alpha) \cdot \log(C_{L}(\theta)), \label{eq:hybrid_cost}
\end{align}
}
where $0 \leq \alpha \leq 1$ is a weighting coefficient. The rationale for including the logarithmic term stems from the behavior of $C_{L}(\theta)$, which approaches zero as the solution improves. The logarithm, $\log(C_{L}(\theta))$, approaches $-\infty$ in this limit, and its derivative magnitude increases significantly as $C_{L}(\theta) \to 0$. This property can help amplify the gradient signal during the later stages of optimization when the cost function value is small, potentially preventing the optimizer from stalling and leading to more stable convergence. This hybrid approach leverages the strengths of both formulations, proving particularly effective for the complex matrices encountered in NEGF.

\subsection{Efficient Decomposition Method for the Cost Function}
A critical step in implementing VQLS on quantum hardware is decomposing operators into forms that can be efficiently measured. Specifically, evaluating our cost functions requires measuring expectation values involving the operators $A$ and $A^\dagger A$. This necessitates expressing these operators as linear combinations of Pauli strings.

Recall the matrix $A$ in the quantum transport equation is $A = (E + i0^+)I - H_0 - D$, where $H_0$ is the base Hamiltonian and $D$ collects diagonal potential energy and self-energy terms. The operator $A^\dagger A$, central to the normalized-residual cost function (Eq.~\ref{eq:cnr_theta}), expands as:
{
\setlength{\abovedisplayskip}{4pt}
\setlength{\belowdisplayskip}{4pt}
\begin{align}
A^\dagger A =& (EI - H_0 - D)^\dagger(EI - H_0 - D) \nonumber \\
=& E^2I - E(H_0 + H_0^\dagger) - E(D + D^\dagger) + H_0^\dagger H_0 \nonumber \\
& + H_0 D + D^\dagger H_0 + D^\dagger D. \label{eq:AdagA_expansion_full}
\end{align}
}

We demonstrate the efficient decomposition of each distinct type of term in this expansion using the shift-operator formalism, which relies on multi-controlled Toffoli gates implementable on certain quantum platforms (e.g., quantum dot arrays\cite{qiScalableMultiqubitIntrinsic2024}).

\subsubsection{Identity and Diagonal Terms}
The $E^2 I$ term requires no decomposition. The diagonal terms $D$, $D^\dagger$, and $D^\dagger D$ stem from the potential profile $E_{sub}$ and self-energies $\Sigma_{1,2,S}$. While any $N \times N$ diagonal matrix generally requires up to $N = 2^n$ Pauli strings, the typically smooth potential profiles in physically realistic semiconductor devices allow $D$ to be approximated accurately using only polynomially many $I/Z$ Pauli strings. We leverage this physical property for efficient decomposition of all diagonal matrix terms.

\subsubsection{Linear $H_0$ Terms}
For the 1D tight-binding lattice with parameter $t_0=\frac{\hbar^2}{2m^*a^2}$, $H_0$ can be expressed using shift operators $S^{(n)} = \sum_{i=0}^{2^n-1}|(i+1) \mod 2^n\rangle \langle i|$ and projector $I_0 = |0 \rangle \langle 0|$ as\cite{sato2021variational}:
{
\setlength{\abovedisplayskip}{4pt}
\setlength{\belowdisplayskip}{4pt}
\begin{align}
H_0 = &t_0[2I^{\otimes n} - I^{\otimes n-1}\otimes X\nonumber \\
& - {S^{(n)}}^\dagger (I^{\otimes n-1}\otimes X) S^{(n)} + {S^{(n)}}^\dagger (I_0^{\otimes n-1}\otimes X) S^{(n)}]. \label{eq:H_0_decomp_shift}
\end{align}
}

This provides a decomposition into measurable terms, as $S^{(n)}$ can be implemented using multi-controlled gates Toffoli gates).
\subsubsection{Cross Terms involving $H_0$ and $D$}
Terms like $H_0 D$ and $D^\dagger H_0$ involve products of $H_0$ components and the diagonal matrix $D$. Consider a representative term like ${S^{(n)}}^\dagger (I^{\otimes n-1}\otimes X) S^{(n)}D$. Using the identity $S^{(n)}D = (S^{(n)} D S^{(n)\dagger}) S^{(n)}$, where $D' = S^{(n)} D S^{(n)\dagger}$ is also a diagonal matrix (representing the shifted potential), the term becomes ${S^{(n)}}^\dagger (I^{\otimes n-1}\otimes X) D' S^{(n)}$. Since $D'$ is diagonal and assumed polynomially decomposable into $I/Z$ strings (due to smoothness), this combined term can be measured by applying the $S^{(n)}$ operations and measuring the expectation value of $(I^{\otimes n-1}\otimes X)$ multiplied by the $I/Z$ Pauli strings from $D'$. Similar arguments apply to all components of $H_0 D$ and $D^\dagger H_0$, indicating they are also decomposable into polynomially many Pauli strings.

\subsubsection{Quadratic $H_0$ Term ($H_0^\dagger H_0$)}
This term requires decomposing the product of $H_0^\dagger$ and $H_0$. Using the shift operator formalism (Eq.~\ref{eq:H_0_decomp_shift}), we have:
{
\setlength{\abovedisplayskip}{4pt}
\setlength{\belowdisplayskip}{4pt}
\begin{align}
H_0^2 =& t_0^2\big[6I^{\otimes n} - 4I^{\otimes n\text{-}1}\otimes X - 4{S^{(n)}}^\dagger (I^{\otimes n\text{-}1}\otimes X) S^{(n)} \nonumber \\
& + 4{S^{(n)}}^\dagger (I_0^{\otimes n\text{-}1}\otimes X) S^{(n)} - {S^{(n)}}^\dagger (I_0^{\otimes n\text{-}1}\otimes I) S^{(n)} \nonumber \\
& + J_1 + J_2 \big], \label{eq:H02_intermediate}
\end{align}
}
where $J_1=I^{\otimes n\text{-}1}\otimes X {S^{(n)}}^\dagger (I^{\otimes n\text{-}1}\otimes X - I_0^{\otimes n\text{-}1}\otimes X) S^{(n)}$ and $J_2={S^{(n)}}^\dagger (I^{\otimes n\text{-}1}\otimes X - I_0^{\otimes n\text{-}1}\otimes X) S^{(n)} (I^{\otimes n\text{-}1}\otimes X)$. Using the recursive property of the shift operator $S^{(n)}$ and identities such as:
{
\setlength{\abovedisplayskip}{4pt}
\setlength{\belowdisplayskip}{4pt}
\begin{align}
{S^{(n)}}^\dagger (I^{\otimes n-1}\otimes X) S^{(n)} &= S^{(n-1)} \otimes |0\rangle\langle1| + {S^{(n-1)}}^\dagger |1\rangle\langle0|, \nonumber \\
{S^{(n)}}^\dagger (I_0^{\otimes n-1}\otimes X) S^{(n)} &= |0\rangle^{\otimes n}\langle 1|^{\otimes n} + |1\rangle^{\otimes n}\langle 0|^{\otimes n},
\end{align}
}
one can show after algebraic manipulation that the sum $J_1 + J_2$ simplifies to:
{
\setlength{\abovedisplayskip}{4pt}
\setlength{\belowdisplayskip}{4pt}
\begin{align}
J_1 + J_2 
=&\big[I^{\otimes n\text{-}2}\otimes X + {S^{(n\text{-}1)}}^\dagger (I^{\otimes n\text{-}2}\otimes X) S^{(n\text{-}1)} \nonumber \\
& - {S^{(n\text{-}1)}}^\dagger (I_0^{\otimes n\text{-}2}\otimes X) S^{(n\text{-}1)}\big]\otimes I.
\end{align}
}
Substituting this back into Eq.~\ref{eq:H02_intermediate}, the full decomposition for $H_0^2$ (or $H_0^\dagger H_0$ when $H_0$ is Hermitian) becomes:
{
\setlength{\abovedisplayskip}{4pt}
\setlength{\belowdisplayskip}{4pt}
\begin{align}
H_0^2 =& t_0^2\big[6I^{\otimes n} - 4I^{\otimes n\text{-}1}\otimes X - 4{S^{(n)}}^\dagger (I^{\otimes n\text{-}1}\otimes X) S^{(n)} \nonumber \\
& + 4{S^{(n)}}^\dagger (I_0^{\otimes n\text{-}1}\otimes X) S^{(n)} - {S^{(n)}}^\dagger (I_0^{\otimes n\text{-}1}\otimes I) S^{(n)} \nonumber \\
& + {S^{(n\text{-}1)}}^\dagger ((I^{\otimes n\text{-}2}-I_0^{\otimes n\text{-}2})\otimes X) S^{(n\text{-}1)} \otimes I \nonumber \\
& + I^{\otimes n\text{-}2}\otimes X \otimes I \big]. \label{eq:H02_final_shift_decomp}
\end{align}
}
Like $H_0$, this decomposition involves terms measurable using circuits implementing the shift operator $S^{(n)}$ (and $S^{(n-1)}$), which can be constructed from multi-controlled Toffoli gates. This detailed decomposition confirms that the $H_0^\dagger H_0$ term can also be expressed as a linear combination of efficiently measurable Pauli strings.

\subsubsection{Implications for Measurement}
The term-by-term analysis confirms that, under the assumption of polynomial decomposability for the diagonal matrix $D$ based on physical smoothness, all constituent operators within $A^\dagger A$ (and thus $A$) can be expressed as a linear combination of polynomially many Pauli strings using the shift-operator formalism. The expectation value of each Pauli string can be estimated to precision $\delta$ using $O(1/\delta^2)$ measurements. Therefore, the overall cost function can be evaluated on a quantum computer with computational resources scaling polynomially in the number of qubits $n = \log(N)$ and polynomially in the desired precision $1/\epsilon$. This efficient decomposition and measurement strategy is fundamental to the VQLS approach, ensuring that the cost function evaluation does not become a bottleneck and preserving the potential for quantum advantage stemming from the logarithmic scaling with the problem dimension $N$ compared to classical methods.

\begin{figure*}[htb!]
    \centering
    \begin{subfigure}[t]{0.237\textwidth}
    \centering
    \includegraphics[width=\linewidth]{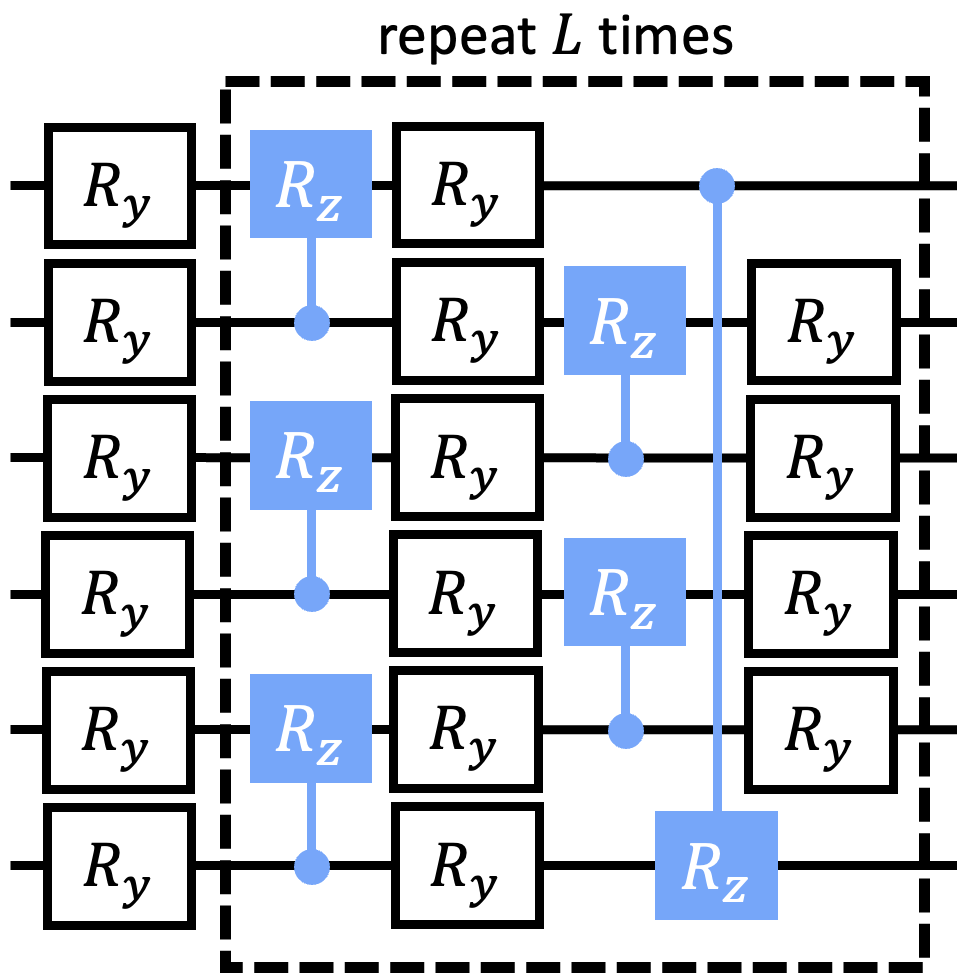}
    \caption{}
    \label{fig:CRZ_circular}
    \end{subfigure}
    \begin{subfigure}[t]{0.230\textwidth}
    \centering
    \includegraphics[width=\linewidth]{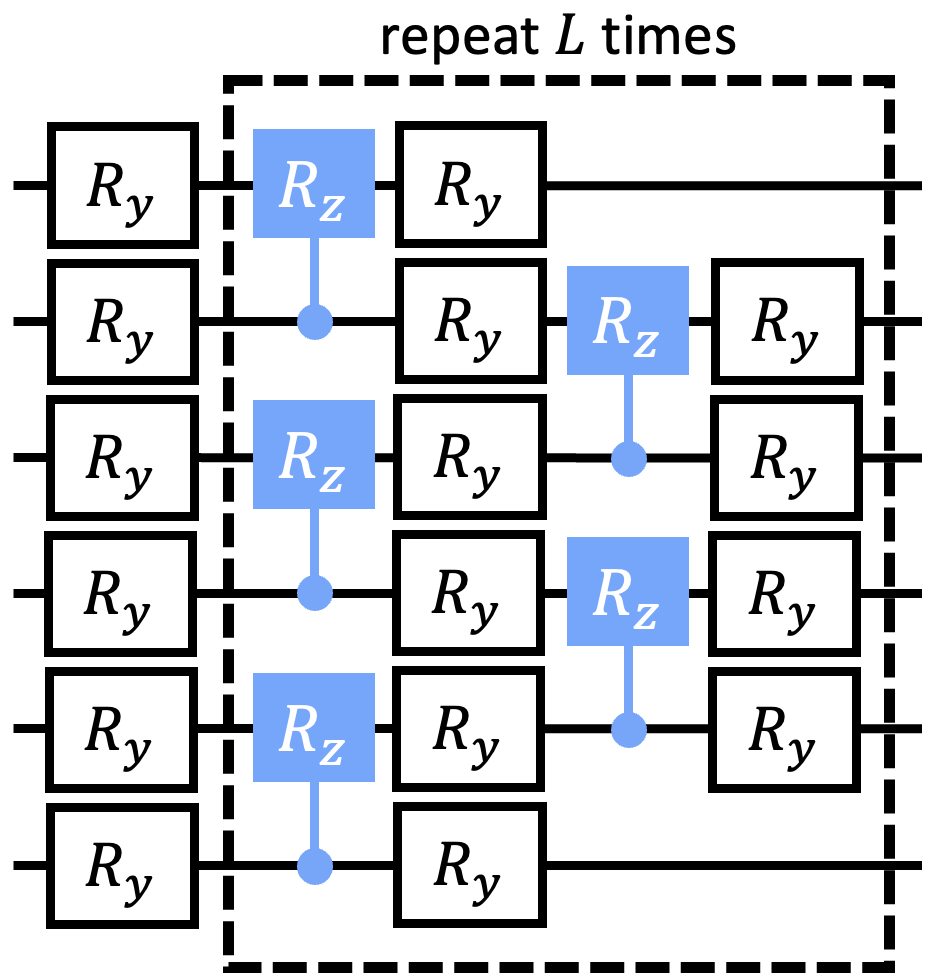}
    \caption{}
    \label{fig:CRZ_linear}
    \end{subfigure}
    \begin{subfigure}[t]{0.205\textwidth}
    \centering
    \includegraphics[width=\linewidth]{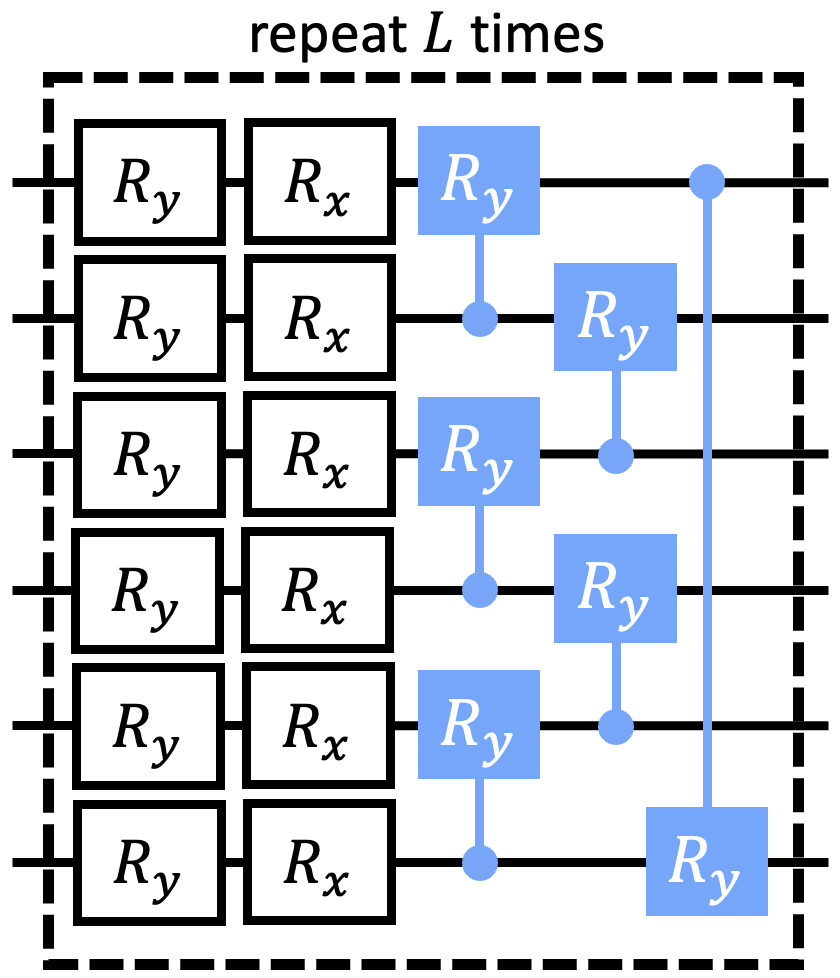}
    \caption{}
    \label{fig:CRY_circular}
    \end{subfigure}
    \begin{subfigure}[t]{0.184\textwidth}
    \centering
    \includegraphics[width=\linewidth]{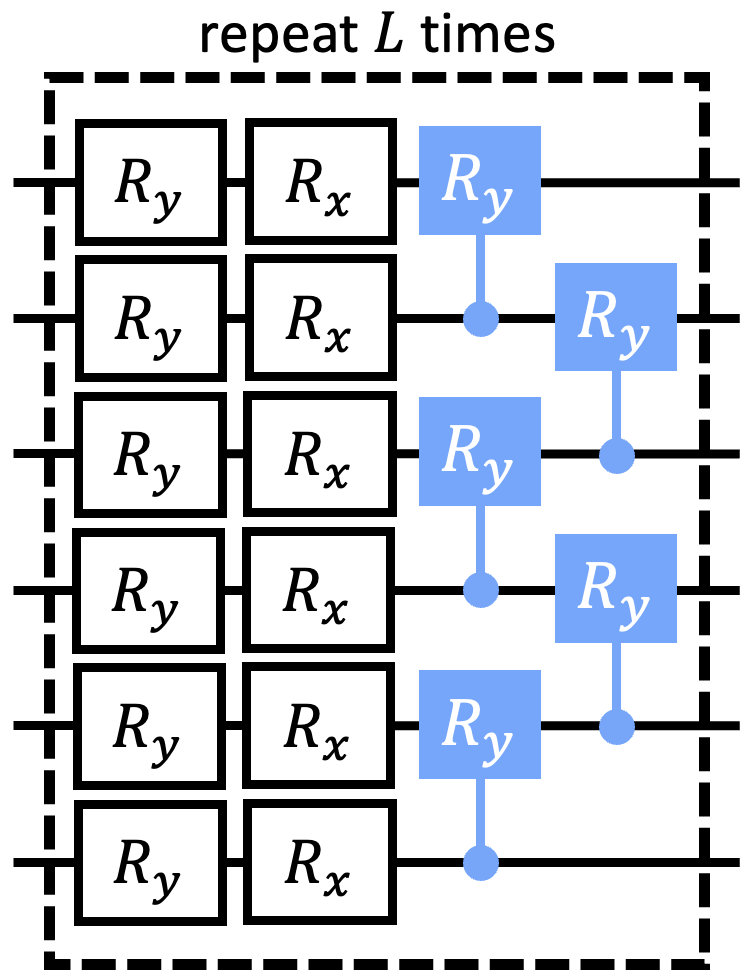}
    \caption{}
    \label{fig:CRY_linear}
    \end{subfigure}
    \caption{Candidate hardware‑efficient ansatz for the NEGF solver. Each circuit block (dashed rectangle) is repeated $L$ times.  White boxes are single‑qubit rotations applied to every qubit; blue boxes mark the parameterized two‑qubit entanglers in a layer. (a) Alternating $CR_z\text{-}R_y$ (circular) – $R_y$ on all qubits followed by $CR_z$ gates. (b) Alternating $CR_z\text{-}R_y$ (linear) – same gates, but entanglement only between nearest neighbours. (c) $R_y\text{-}R_x\text{-}CR_y$ (circular) – two single‑qubit rotations per qubit before circular $CR_y$ entanglers. (d) $R_y\text{-}R_x\text{-}CR_y$ (linear) – identical gate set with linear entanglement.}
    \label{fig:ansatz}
    \vspace{-15pt}
\end{figure*}

\subsection{Ansatz Design}
The design of an effective quantum ansatz, the parameterized quantum circuit $U(\theta)$ used to represent the solution state $|\psi(\theta)\rangle = U(\theta)|0\rangle$, is also crucial for the success of VQLS. The ansatz must be sufficiently expressive to capture the true solution within the reachable state space, yet remain trainable on NISQ devices, avoiding issues like barren plateaus and excessive gate depths.

Hardware-efficient ansatz, constructed using native or easily implementable gates on a given quantum processor, are commonly employed. Typical structures involve alternating layers of single-qubit rotations and two-qubit entangling gates. However, many standard hardware-efficient ansatz, such as those using only $R_y$ single-qubit rotations and CNOT or controlled-$Z$ gates, are primarily designed for problems where the solution is expected to be real-valued.

The NEGF formalism, however, inherently involves complex-valued Green's functions, meaning the target solution vector $G_i(E)$ has complex components. Therefore, the chosen ansatz must possess adequate expressibility to represent both the real and imaginary parts of the solution state effectively. To adapt hardware-efficient structures for this complex-valued problem, we consider two main strategies:

\begin{itemize}
    \item \textbf{Parameterized Entanglement:} Modify the standard two-qubit entangling gate to introduce complex phases directly. One approach is to use single-qubit $R_y$ gates followed by a parameterized controlled-$R_z$ rotation ($CR_z$), for entanglement. This creates an alternating layered structure where the entanglement operation itself contributes to the complex phase generation.
    \item \textbf{Expanded Single-Qubit Rotations:} Expand the single-qubit rotation layer to manipulate both amplitude and phase. This can be achieved by using two types of single-qubit rotations per qubit in each layer, for example, $R_y$ potentially combined with $R_x$ (or $R_z$), before the entanglement block. This allows for independent control over different components of the single-qubit states.
\end{itemize}

In addition to the gate choice, the entanglement pattern plays a significant role. We consider two common topologies for connecting the $n$ qubits:
\begin{itemize}
    \item \textbf{Linear Entanglement:} Entangling gates are applied only between adjacent qubit pairs $(j, j+1)$ for $j=0, \dots, n-2$.
    \item \textbf{Circular Entanglement:} Entangling gates are applied between adjacent pairs $(j, (j+1) \mod n)$ for $j=0, \dots, n-1$, effectively connecting the first and last qubits.
\end{itemize}

We propose and investigate four candidate ansatz structures for solving the complex-valued NEGF problem, as shown in \cref{fig:ansatz}, combining the gate strategies and entanglement topologies discussed above. These include circular and linear variants of both the alternating $CR_z\text{-}R_y$ configuration and the $R_y\text{-}R_x\text{-}CR_y$ configuration.

The performance of these candidate ansatz, including their expressibility and trainability for the NEGF problem, will be evaluated numerically in Sec.~\ref{sec: results}. The goal is to identify an ansatz structure that balances representation capability for complex states with the practical constraints of NISQ hardware.

\section{Results and Discussion} \label{sec: results}
In this section, we present numerical experiments to validate our proposed VQLS approach for solving quantum transport equations. We focus on simulating quantum transport with a 10 nm nanosheet Field-Effect Transistor (FET)\cite{liuMonolayer$rmMoS_2$2013} to demonstrate the practical applicability and performance of our methods. We evaluate the key components of our VQLS framework, including the effectiveness of the proposed cost functions, the performance of different ansatz designs tailored for complex-valued problems, and the advantages offered by quantum parallelism for NEGF computations.

\subsection{Cost Function Evaluation}
The choice of cost function significantly impacts the convergence and the accuracy of the final solution obtained by VQLS. We systematically evaluated the performance of four different cost functions discussed in Section~\ref{sec: approach}: the traditional global ($C_{G}$) and local ($C_{L}$) cost functions, the proposed normalized-residual cost function ($C_{NR}$), and our proposed hybrid cost function ($C$) which combines $C_{NR}$ and $\log(C_{L})$.

For these experiments, we simulated the 1-sheet nanosheet FET problem discretized into 32 grid points, requiring $n=5$ qubits. The simulations were performed for a single injection energy of $E = -0.044$ eV. We employed the $CR_z\text{-}R_y$ ansatz with circular entanglement in Fig.~\ref{fig:CRZ_circular} with the number of layers $L$ varied from 3 to 6. Parameters were initialized using the beta distribution $\text{Beta}(0.5, 0.5) \times \pi$. The BFGS algorithm served as the classical optimizer. For each configuration (cost function and layer number), 10 independent optimization runs were performed to gather statistics. For the hybrid cost function (Eq.~\ref{eq:hybrid_cost}), the weight parameter $\alpha$ was initially set to 0.7 unless otherwise specified.

\begin{table}[htbp]
\centering
\caption{MSE statistics comparing VQLS solutions for different cost functions and ansatz layers ($L=3$ to $6$).}
\label{tab:mse_results}
\begin{tabular}{@{}c@{\hskip 2pt}cccc@{}}
\hline
\textbf{Cost Function} & \textbf{Layers} & \textbf{Mean} & \textbf{Median} & \textbf{Minimum} \\
\hline
\multirow{4}{*}{Global Cost} 
& 3 & 316.268 & 418.470 & 5.388 \\
& 4 & 209.516 & 138.533 & 14.670 \\
& 5 & 301.055 & 287.445 & 0.290 \\
& 6 & 250.381 & 316.457 & 1.324 \\
\hline
\multirow{4}{*}{Local Cost} 
& 3 & 274.991 & 345.924 & 21.023 \\
& 4 & 190.872 & 113.881 & 0.385 \\
& 5 & 227.191 & 251.098 & 40.067 \\
& 6 & 236.058 & 178.376 & 9.206 \\
\hline
\multirow{4}{*}{Normalized Residual} 
& 3 & 2.753 & 1.834 & 1.468 \\
& 4 & 27.517 & 1.270 & 0.545 \\
& 5 & 0.544 & 0.633 & 0.134 \\
& 6 & 0.423 & 0.356 & 0.164 \\
\hline
\multirow{4}{*}{\textbf{Hybrid Cost}}
& 3 & \textbf{1.203} & \textbf{1.102} & \textbf{0.748} \\
& 4 & \textbf{0.665} & \textbf{0.657} & \textbf{0.322} \\
& 5 & \textbf{0.246} & \textbf{0.204} & \textbf{0.101} \\
& 6 & \textbf{0.055} & \textbf{0.021} & \textbf{5.549e-05} \\
\hline
\end{tabular}
\vspace{-8pt}
\end{table}

Table~\ref{tab:mse_results} summarizes the Mean Squared Error (MSE) statistics from these runs. A clear difference in performance is observed. The traditional global and local cost functions exhibit very unstable training behavior; while the minimum achieved MSE can occasionally be relatively low (e.g., 0.290 for global cost function at $L=5$), the mean and median MSE values remain extremely high across all layers, indicating frequent convergence to poor solutions. Furthermore, these cost functions appear susceptible to issues like barren plateaus, as evidenced by the local cost function where the mean, median, and minimum MSE all increase when going from $L=4$ to $L=5$ layers.

In contrast, the normalized-residual cost function achieves significantly better and more consistent results, with mean and median MSE values orders of magnitude lower than the global and local approaches for $L=3$, $5$, and $6$. The hybrid cost function demonstrates the best performance overall. It consistently yields the lowest mean, median, and minimum MSE values across all layer depths, showcasing its ability to further improve solution accuracy and enhance the stability of the training process compared to the already effective normalized-residual approach.

\begin{figure}[ht!]
    \centering
    \includegraphics[width=\linewidth]{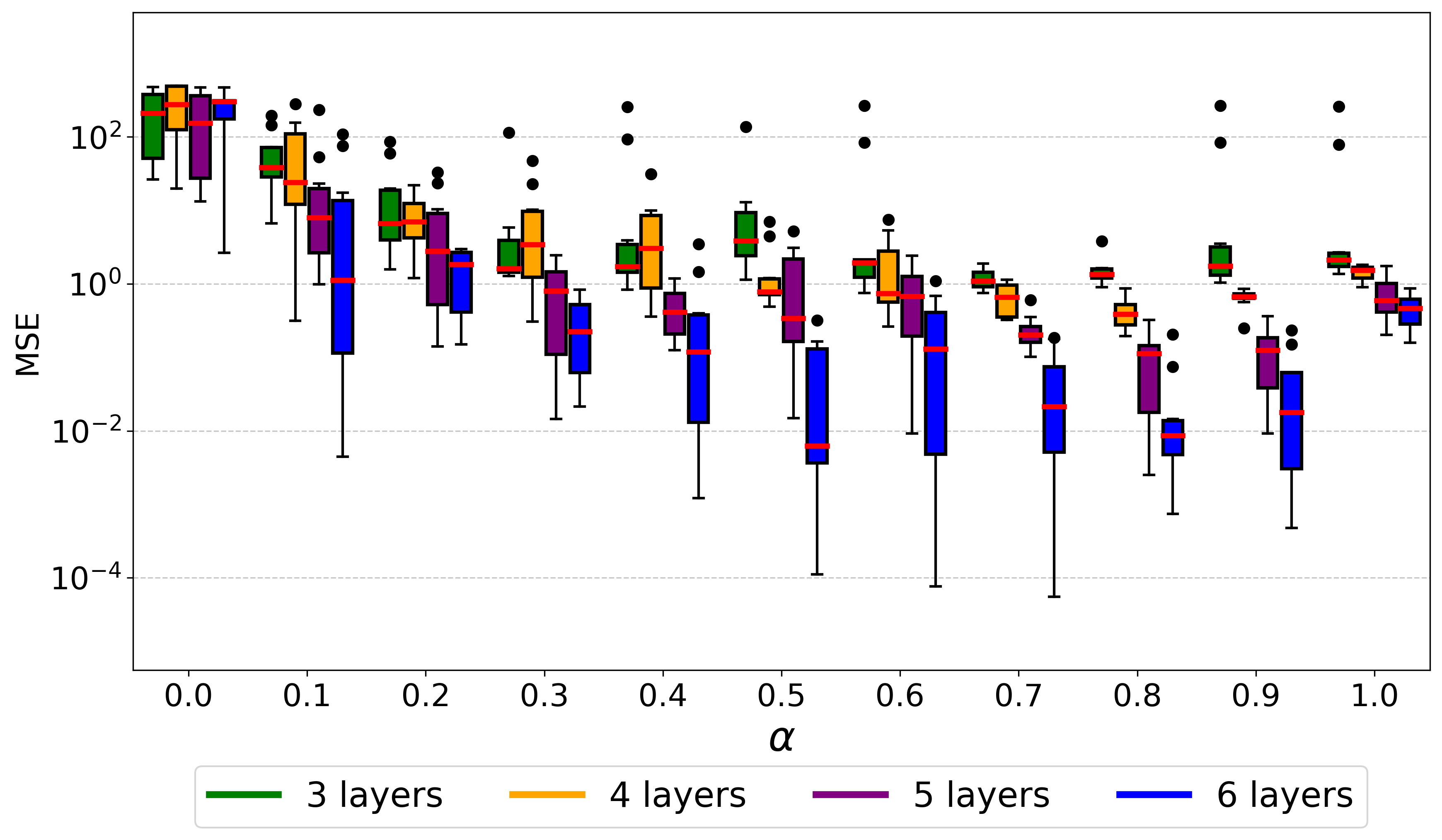}
    \caption{MSE values for different weight values, $\alpha$, in the hybrid cost function (Eq.~\ref{eq:hybrid_cost}) for the 5-qubit nanosheet FET problem at $E = -0.044$ eV, evaluated for $L=4$, $5$, and $6$ layers using the $CR_z\text{-}R_y$ ansatz with circular entanglement. The horizontal axis represents $\alpha$ values, ranging from 0 (log local cost function) to 1 (normalized-residual cost function).}
    \label{fig:cost_weight_comp}
    \vspace{-15pt}
\end{figure}

To further investigate the hybrid cost function, we explored the impact of the weighting parameter $\alpha$ on the solution quality. Figure~\ref{fig:cost_weight_comp} shows the resulting MSE as $\alpha$ is varied from 0 (purely log-local cost) to 1 (purely normalized-residual cost) for different numbers of ansatz layers. The analysis reveals that the normalized-residual component ($\alpha = 1$) generally leads to better solutions than the log-local component ($\alpha = 0$) for this problem. However, incorporating the logarithmic term $\log(C_{L})$ with a non-zero weight $(1-\alpha)$ often improves performance further, likely by enhancing gradient magnitudes during optimization. Consistently across different layer depths, the optimal solution quality is achieved with $\alpha$ values typically ranging between 0.7 and 0.9. This confirms the effectiveness of the hybrid approach in balancing direct residual minimization with the beneficial optimization properties of the logarithmic local cost term for NEGF simulations.

\subsection{Ansatz Design and Parameter Initialization}
Having established the effectiveness of the hybrid cost function, we now evaluate the performance of the four candidate hardware-efficient ansatz structures shown in \cref{fig:ansatz}, designed to handle the complex-valued nature of the NEGF problem. The goal is to identify an ansatz that provides a good balance between expressibility and practical efficiency for simulating the nanosheet FET.

The experimental setup remains consistent with the previous: a 5-qubit system (32 grid points), injection energy $E = -0.044$ eV, Beta initialization ($\text{Beta}(0.5, 0.5) \times \pi$), and the BFGS optimizer over 10 independent runs per configuration. We specifically use the hybrid cost function with the optimized weight $\alpha = 0.7$. The number of layers $L$ in each ansatz is varied to assess performance dependency on circuit depth.

\begin{figure}[h]
    \centering
    \includegraphics[width=0.85\linewidth]{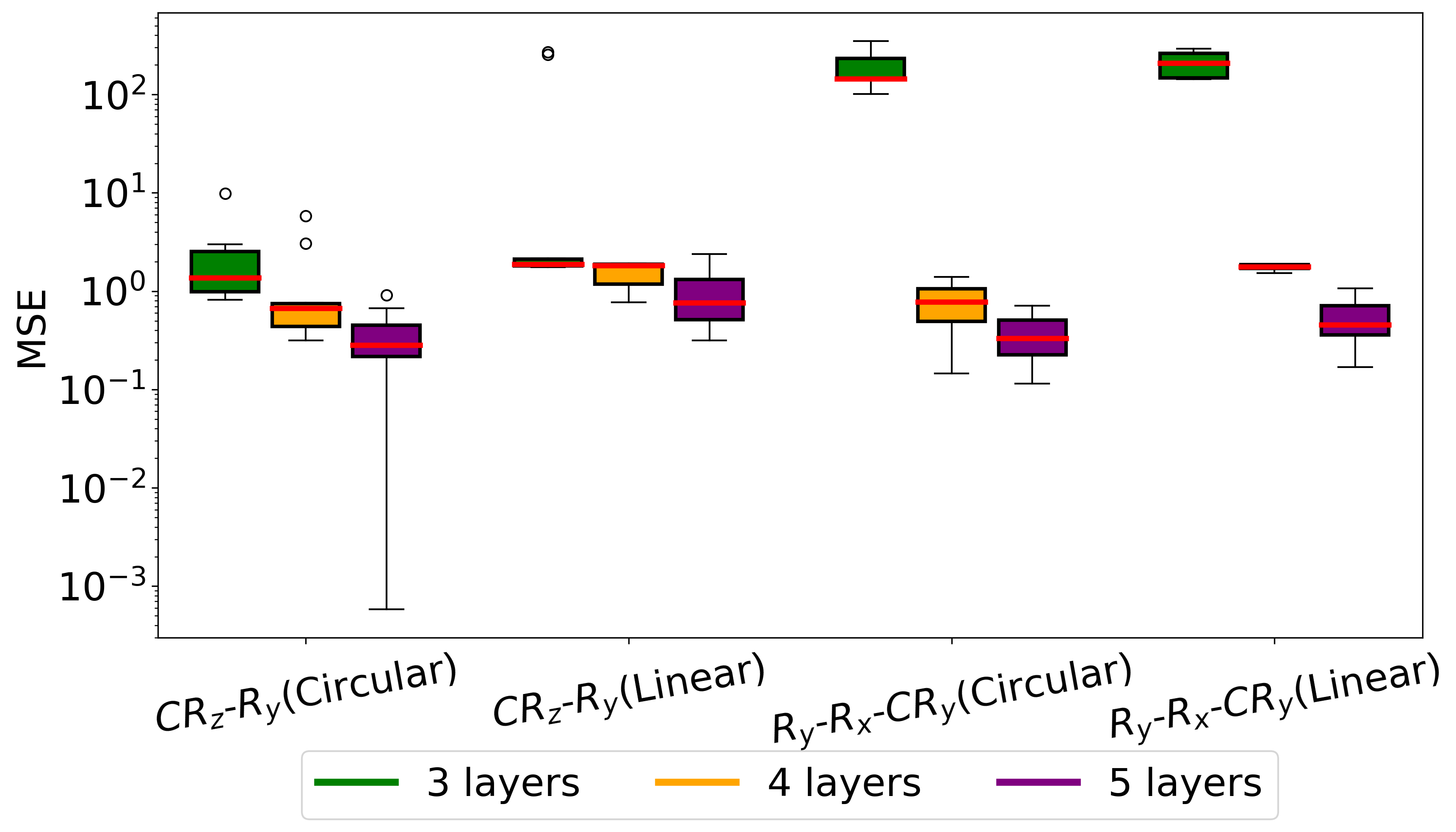}
    \caption{Box plot comparing the final MSE distribution over 10 runs for the four candidate ansatz designs across different numbers of layers ($L$). Results shown for the 5-qubit nanosheet FET simulation ($E = -0.044$ eV) using the hybrid cost function ($\alpha=0.7$). Whiskers indicate variability, boxes show interquartile range, and red lines mark the median.}
    \label{fig:ansatz_comp}
    \vspace{-10pt}
\end{figure}

Figure~\ref{fig:ansatz_comp} presents box plots summarizing the distribution of the MSE achieved after optimization for each of the four candidate ansatz across different numbers of layers ($L$). The results indicate two clear trends. First, comparing entanglement topologies, the circular versions consistently tend to achieve lower median and minimum MSE compared to their linear counterparts. This suggests the circular topology better captures the problem's structure and aids the optimization process. Second, comparing the gate strategies, the alternating $CR_z\text{-}R_y$ structure generally outperforms the layered $R_y-R_x-CR_y$ structure, showing lower errors especially for a limited number of layers ($L=3$).
\begin{figure}[h]
    \centering
    \includegraphics[width=0.7\linewidth]{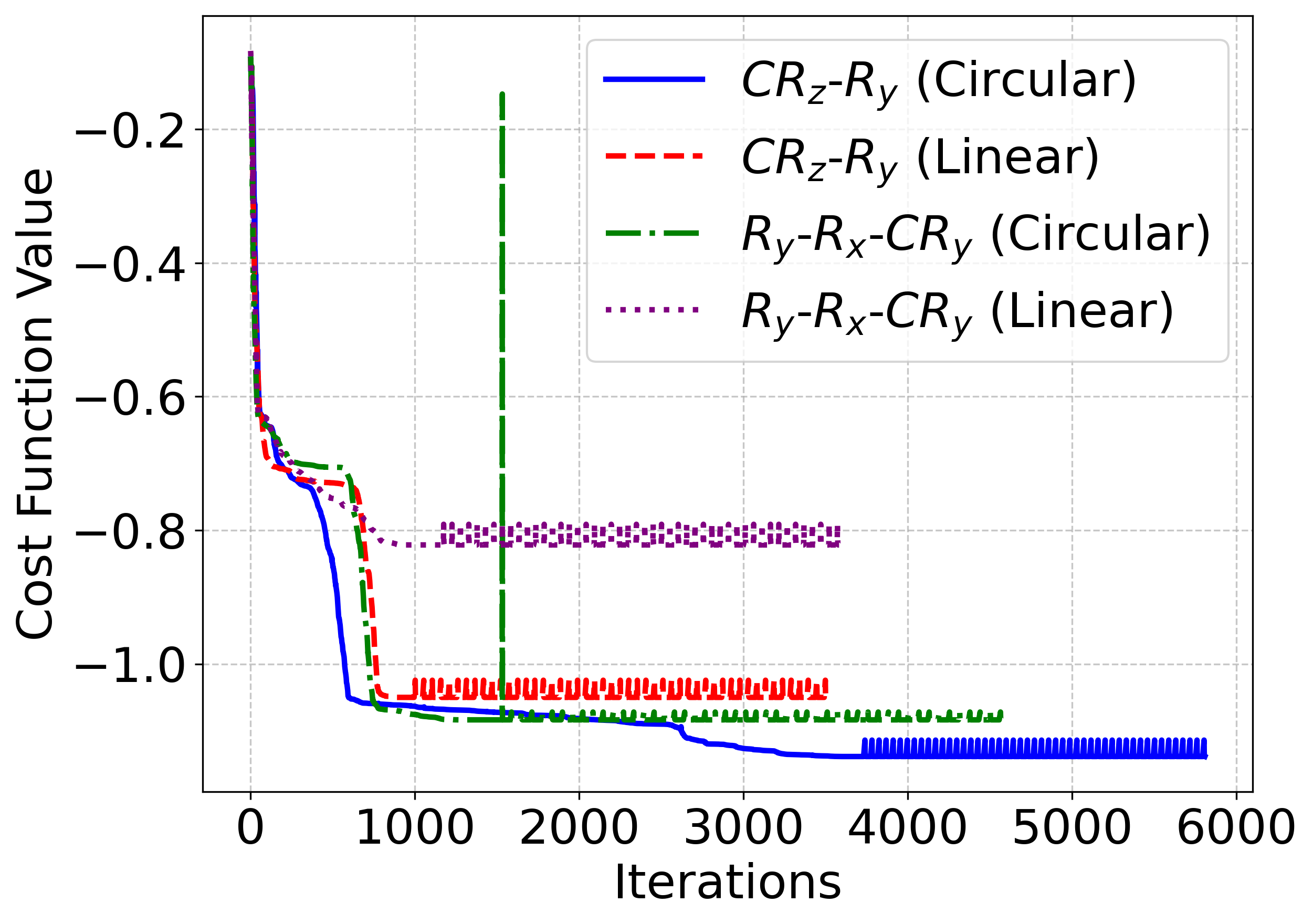}
    \caption{Evolution of the hybrid cost function value versus optimization iteration for the four candidate ansatz designs with $L=3$ layers. The plot illustrates the convergence speed and final achieved cost value for each ansatz structure.}
    \label{fig:cost_his_comp}
    \vspace{-10pt}
\end{figure}

To assess the convergence behavior, \cref{fig:cost_his_comp} plots the evolution of the hybrid cost function value as a function of the optimization iterations for ansatz with $L=3$ layers. This comparison highlights differences in convergence speed and the final cost value achieved. The $CR_z\text{-}R_y$ ansatz with circular entanglement not only converges significantly faster but also reaches a consistently lower final cost function value compared to the other three candidates for this circuit depth.


Considering both the final solution accuracy and the convergence efficiency, the $CR_z\text{-}R_y$ ansatz with circular entanglement emerges as the most promising choice among the tested candidates for this VQLS application to the NEGF problem. It provides a favorable combination of expressibility for complex-valued states and efficient trainability under the chosen conditions.

\subsection{Quantum Parallelism in NEGF Computation}
A significant potential advantage of employing VQLS for NEGF simulations lies in leveraging quantum properties to perform computations across multiple energy values $E$ simultaneously. While classical approaches require solving the NEGF equation sequentially for each energy point, a quantum approach can encode the solutions for all energies within a single, larger quantum state. This parallelism arises fundamentally from the nature of the quantum state prepared by the ansatz circuit $U(\theta)$. An $n$-qubit state $|\psi(\theta)\rangle$ exists as a superposition within a Hilbert space of dimension $2^n$, allowing it to represent an exponential number of configurations concurrently. By mapping different energy points to different sectors of this large Hilbert space, we can solve for all energies in parallel within one VQLS optimization, unlike classical methods that scale linearly with the number of energy points $N_E$.

To implement this, we construct a larger, combined linear system encompassing all $N_E$ energy levels of interest. The overall system matrix $\mathbf{A}$ becomes block-diagonal, with each block corresponding to the matrix $A(E_i)$ for a specific energy $E_i$. Similarly, the right-hand side vector $\mathbf{b}$ is constructed by concatenating the relevant vectors for each energy block. It's important to note that for simulating ```tic transport in transistors, as demonstrated here, the source vectors $b_k$ (e.g., $b_{source}$ representing injection from the source contact, typically the first column of the identity matrix, and $b_{drain}$ for the drain contact, typically the last column) are energy-independent. Therefore, for calculating quantities like LDOS and transmission coefficients which rely on these specific columns of the Green's function, $\mathbf{b}$ is constructed by repeating $[b_{source}, b_{drain}]^T$ for each energy block.

For our demonstration using the nanosheet FET model with $N_r=32$ spatial grid points (requiring $\log(N_r) = 5$ qubits for a single energy), parallelizing across $N_E=32$ energy points and considering both source ($k=1$) and drain ($k=N_r$) injection vectors requires representing a total system size of $N_r \times N_E \times 2 = 32 \times 32 \times 2 = 2048$. This combined system can be encoded using $n = \lceil\log(2048)\rceil = 11$ qubits. We solved this combined system using our optimized VQLS framework (Hybrid cost function with $\alpha = 0.7$ and the $CR_z\text{-}R_y$ circular ansatz).

\begin{figure}[htb!]
    \centering
    \captionsetup[subfigure]{aboveskip=-1pt,belowskip=-1pt}
    \begin{subfigure}[t]{0.23\textwidth}
    \centering
    \includegraphics[width=\linewidth]{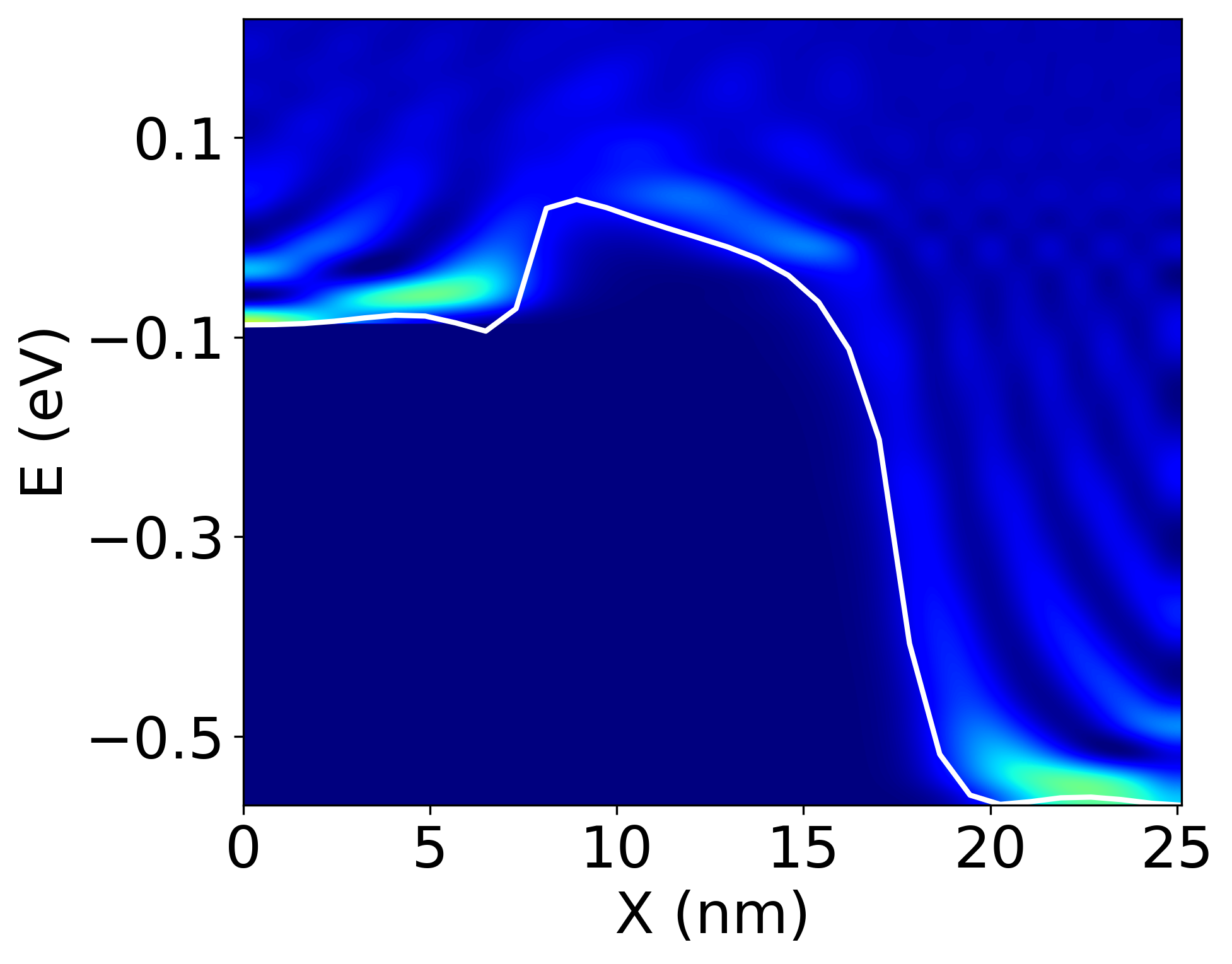}
    \caption{}
    \label{fig:LDOSQ}
    \end{subfigure}
    \begin{subfigure}[t]{0.23\textwidth}
    \centering
    \includegraphics[width=\linewidth]{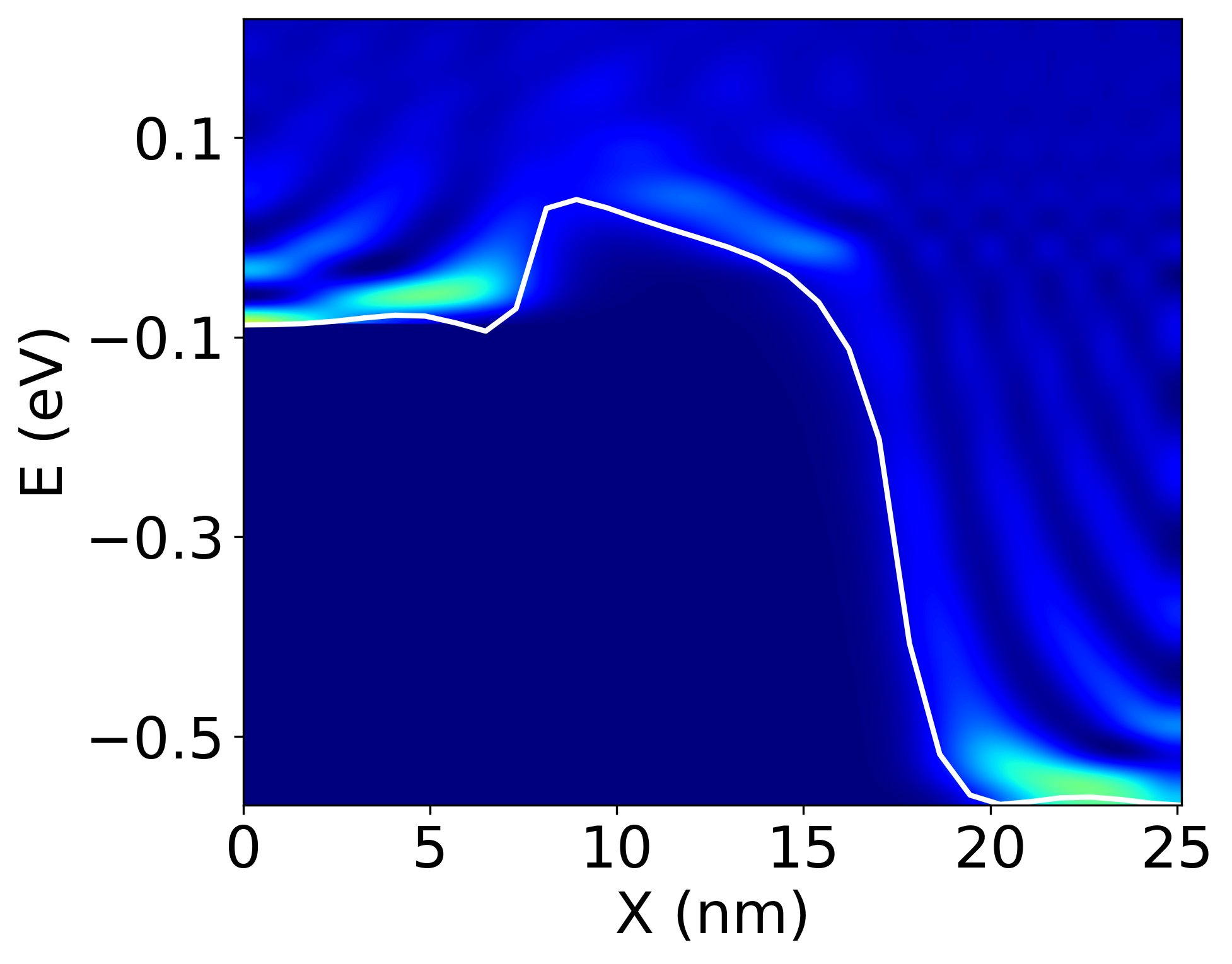}
    \caption{}
    \label{fig:LDOSC}
    \end{subfigure}
    
    \begin{subfigure}[t]{0.23\textwidth}
    \centering
    \includegraphics[width=\linewidth]{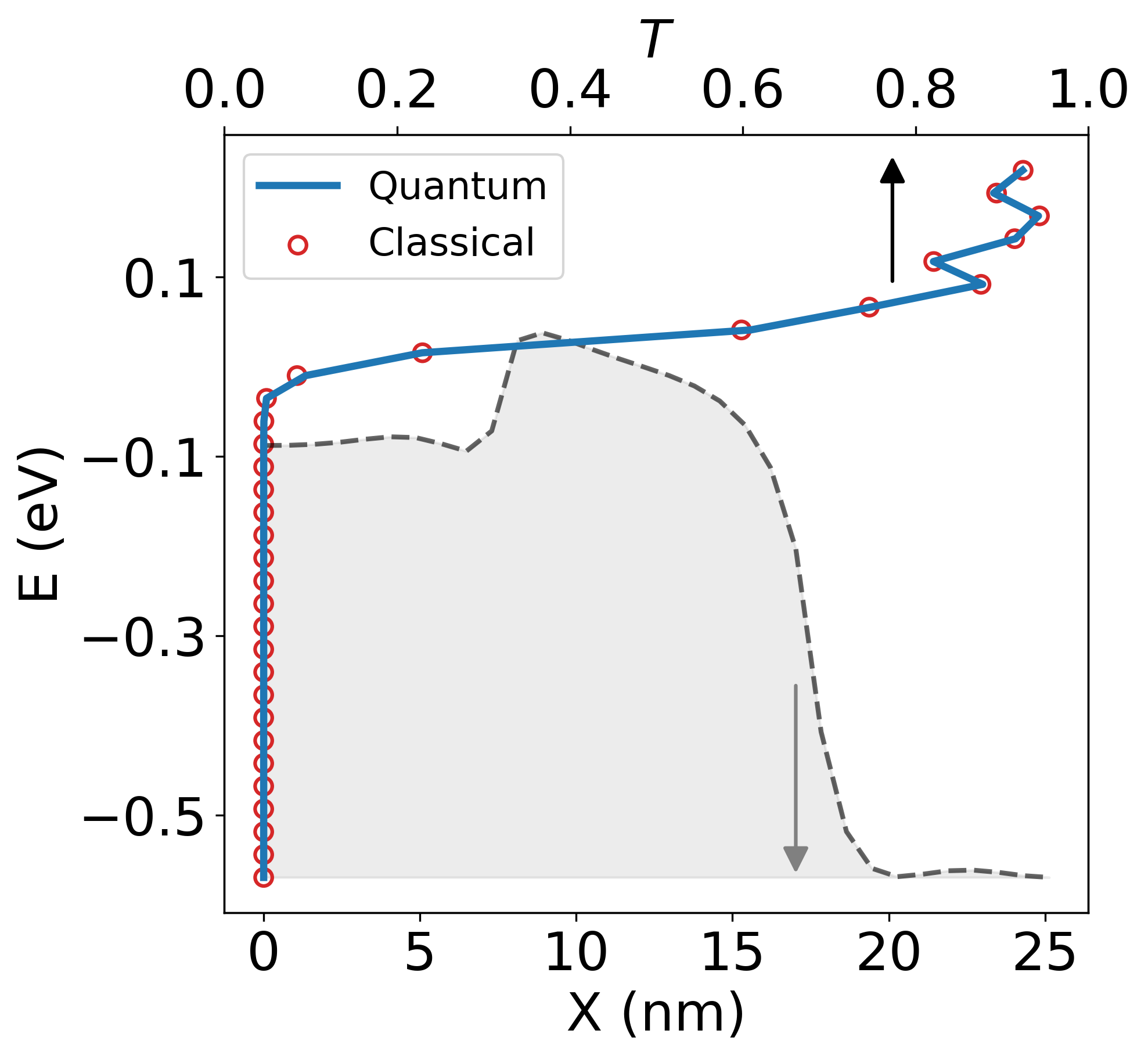}
    \caption{}
    \label{fig:Trans}
    \end{subfigure}
    \begin{subfigure}[t]{0.23\textwidth}
    \centering
    \includegraphics[width=\linewidth]{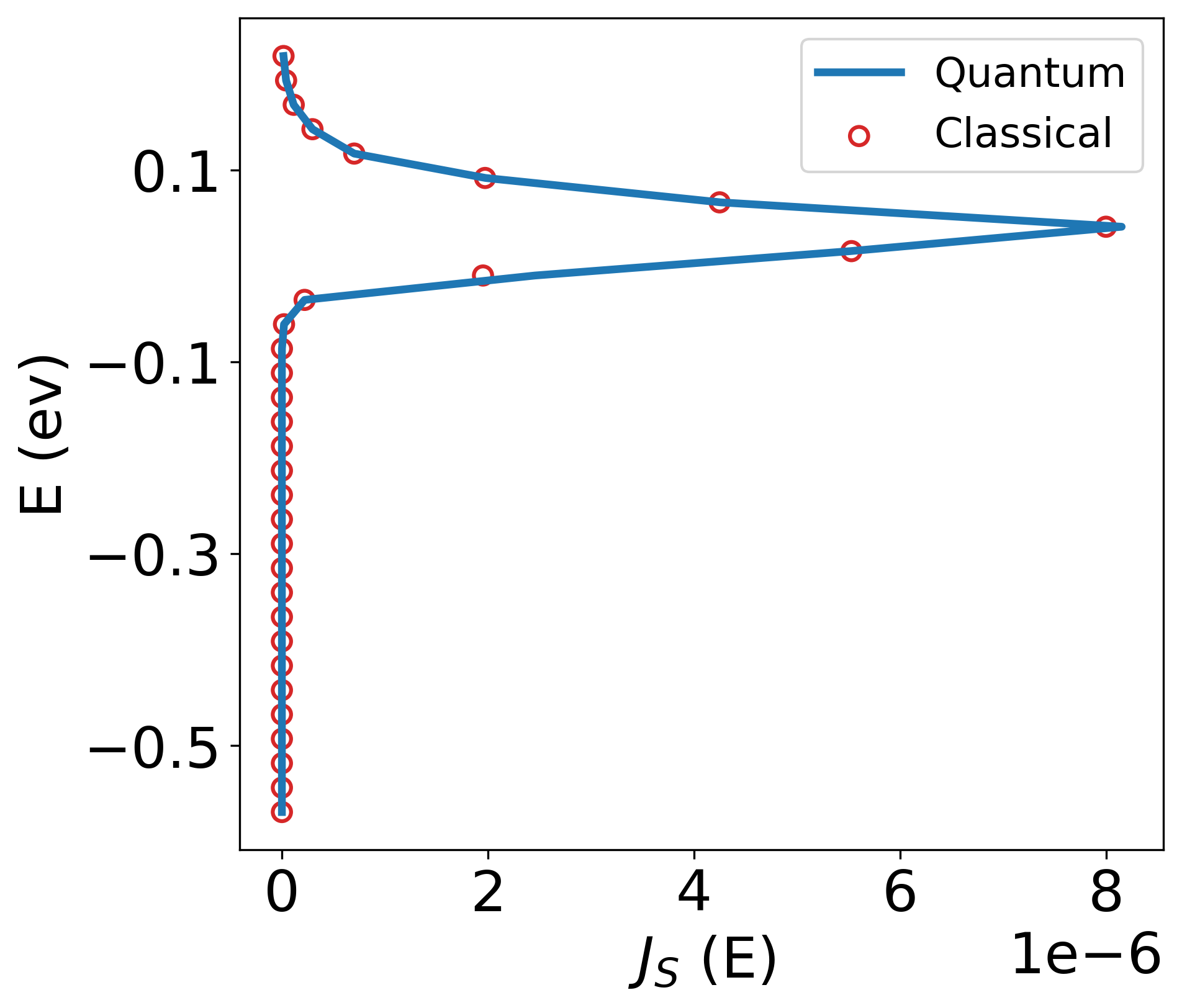}
    \caption{}
    \label{fig:Current}
    \end{subfigure}
    \caption{Comparison between quantum and classical solutions: (a) and (b) show LDOS from quantum and classical approaches; (c) compares transmission coefficients as functions of energy; (d) compares source current densities.}
    \label{fig:pararesult}
    \vspace{-10pt}
\end{figure}

Figure~\ref{fig:pararesult} compares key physical quantities extracted from the parallel VQLS solution against classical NEGF results. The excellent agreement observed in the Local Density of States (LDOS) contour plots (Fig.~\ref{fig:LDOSQ} vs. \ref{fig:LDOSC}), the transmission coefficients $T(E)$ (Fig.~\ref{fig:Trans}), and the energy-resolved current spectra (Fig.~\ref{fig:Current}) validates the accuracy and viability of our parallel quantum approach.

This quantum parallelism provides significant scaling advantages. For instance, parallelizing across 32 energy grid points classically requires approximately 32 times more computing resources compared to a single energy point simulation. In contrast, our quantum approach achieves this by increasing the qubit count from $\log(N_r)$ to $\log(N_r \times N_E \times 2)$, requiring only $11 - 5 = 6$ additional qubits in this example. Similarly, handling 1024 energy points ($N_E = 1024$) would demand roughly 1024 times more classical resources but only $\log(1024 \times 2) = 11$ additional qubits quantum mechanically. This logarithmic scaling in qubit requirement with respect to $N_E$, compared to the linear scaling of classical resources, represents a substantial potential advantage for realistic, large-scale device simulations that necessitate fine energy discretization.

\section{Conclusion} \label{sec: conclusion}
Simulating quantum transport plays a critical role in designing nanoscale semiconductor devices, but classical approaches face significant computational scaling challenges. This work developed a VQLS framework specifically tailored to address the complexities of NEGF simulations, particularly the complex-valued, non-symmetric matrices involved. We propose a hybrid cost function designed for improved convergence and stability with such matrices, along with an efficient decomposition strategy for the cost function.

Our numerical results, which simulate a nanoscale nanosheet FET, validate the proposed VQLS approach with high accuracy. We demonstrated the significant advantage of quantum parallelism inherent in the VQLS state representation, enabling simultaneous calculation across multiple energy levels. This leads to logarithmic scaling ($O(\log(N_E))$) in qubit requirements with the number of energy points $N_E$, contrasting sharply with the linear scaling ($O(N_E)$) of classical computational resources. This work establishes VQLS as a viable tool for quantum transport simulation in semiconductors, paving the way for potentially intractable high-fidelity device simulations on future quantum computers and accelerating the development of next-generation electronics.

\bibliographystyle{IEEEtran}
\bibliography{ref}  

\end{document}